\documentclass{PoS}%
\usepackage{amsmath}
\usepackage{amsfonts}
\usepackage{amssymb}
\usepackage{graphicx}%
\providecommand{\U}[1]{\protect\rule{.1in}{.1in}}
\title{%
\vspace{-2cm}%
{\small
\begin{flushright} \begin{minipage}{4cm}
KEK preprint 2013-65 \\
CHIBA-EP-204
\end{minipage}\end{flushright}}
\vspace{1cm} %
Non-Abelian dual Meissner effect and confinement/deconfinement phase transition in SU(3) Yang-Mills theory%
}

\ShortTitle{Non-Abelian dual Meissner effect and confinement/deconfinement phase transition ..... }

\author{\speaker{Akihiro Shibata} \\%
Computing Research Center, High Energy Accelerator Research Organization (KEK) \& \\
Graduate University for Advanced Studies (Sokendai), Tsukuba 305-0801, Japan \\
E-mail: \email{akihiro.shibata@kek.jp}
}
\author{Kei-Ichi Kondo \\
Department of Physics, Graduate School of Science, Chiba University, Chiba 263-8522, Japan \\
E-mail: \email{kondok@faculty.chiba-u.jp}
}
\author{Seikou Kato \\
Fukui National College of Technology, Sabae, Fukui 916-8507, Japan \\
E-mail: \email{skato@fukui-nct.ac.jp}
}
\author{Toru Shinohara \\
Department of Physics, Graduate School of Science, Chiba University, Chiba 263-8522, Japan \\
E-mail: \email{sinohara@graduate.chiba-u.jp}
}

\abstract{%
The dual superconductivity is a promising mechanism for quark confinement. 
We proposed the non-Abelian dual superconductivity picture for SU(3) Yang-Mills theory, 
and demonstrated the restricted field dominance (called conventionally "Abelian" dominance), 
and non-Abelian magnetic monopole dominance in the string tension. In the last conference, 
we have demonstrated by measuring the chromoelectric flux that the non-Abelian dual Meissner effect
exists and determined that the dual superconductivity for SU(3) case is of type I, which is in sharp contrast 
to the SU(2) case: the border of type I and type II. 

In this talk, we focus on the confinement/deconfinemen phase transition 
and the non-Abelian dual superconductivity at finite temperature: 
We measure the chromoelectric flux between a pair of 
static quark and antiquark at finite temperature, and investigate its relevance 
to the phase transition and the non-Abelian dual Meissner effect.
}

\FullConference{31st International Symposium on Lattice Field Theory LATTICE 2013\\
		 July 29 -- August 3, 2013\\
		 Mainz, Germany}

\begin{document}
\section{Introduction}

Quark confinement follows from the area law of the Wilson loop average. The
dual superconductivity is the promising mechanism for the quark confinement
\cite{dualSC}. Based on the Abelian projection, there have been many numerical
analyses to show evidences such as Abelian dominance \cite{Suzuki90}, Abelian
magnetic monopole dominance \cite{stack94Shiba}, and center vortex dominance
\cite{greensite} in the string tension. However, these results are obtained
only in special gauges such as the maximal Abelian (MA) gauge and the
Laplacian Abelian gauge, and the Abelian projection itself breaks the gauge
symmetry as well as color symmetry (global symmetry).

We have presented the lattice version of a new formulation of $SU(N)$
Yang-Mills (YM) theory\cite{KSM05}, that gives the decomposition of the gauge
link variable $U_{x,\mu}=X_{x,\mu}V_{x,\mu}$ , which is suited for extracting
the dominant mode, $V_{x,\mu}$, for quark confinement in the gauge independent
way. In the case of the $SU(2)$ YM theory, the decomposition of the gauge link
variable is given by a compact representation of Cho-Duan-Ge-Faddeev-Niemi
(CDGFN) decomposition \cite{CFNS-C} on a lattice \cite{ref:NLCVsu2}%
\cite{ref:NLCVsu2-2}\cite{kato:lattice2009}. For the $SU(N)$ YM theory, the
new formula for the decomposition of the gauge link variable is constructed as
an extension of the $SU(2)$ case.

To the Wilson loop in the fundamental representation, we have apply the
minimal option. The minimal option is obtained for the stability group of
$\tilde{H}=U(2)\cong SU(2)\times U(1)$, which is suitable for the Wilson loop
in the fundamental representation. This fact is derived from the non-Abelian
Stokes'\ theorem \cite{KondoNAST}. Then, we have demonstrated the gauge
independent (invariant) restricted $V$-field dominance , (or conventionally
called Abelian\ dominance): the decomposed $V$-field (restricted field)
reproduced the string tension of original YM field $(\sigma_{V}/\sigma
_{full}=93\pm16\%)$, and the gauge independent non-Abelian magnetic monopole
dominance: the string tension was reproduced by only the (non-Abelian)
magnetic monopole part extracted from the restricted field, ($\sigma
_{mon}/\sigma_{V}=94\pm9\%$) \cite{SCGTKKS08L}\cite{lattice2008}%
\cite{lattice2009}\cite{lattice2010}\cite{abeliandomSU(3)}.

To establish the dual superconductivity picture, we must also show the
magnetic monopoles play the dominant role in quark confinement. The dual
Meissner effect in Yang-Mills theory must be examined by measuring the
distribution of chromoelectric field strength or chromo flux as well as
magnetic monopole currents created by a static quark-antiquark pair
\cite{lattice2012}. In $SU(2)$\ case, the extracted field corresponding to the
stability group $\tilde{H}=U(1)$ shows the dual Meissner effect, which is a
gauge invariant version of the Abelian projection in MA gauge. In the SU(3)
case, there are many works on chromo flux for the Yang-Mills field by using
Wilson line/loop operator, e.g., \cite{Cardaci2011}\cite{Cardso}%
\cite{CeaCosmail2012}. However, there is no direct measurement of the dual
Meissner effect in the gauge independent (invariant) way, except for several
studies based on the Abelian projection, e.g., \cite{flusx:AP}. At the last
conference, we have demonstrated the non-Abelian dual Meissner
effect\cite{lattice2012}. By applying our new formulation to the $SU(3)$ YM
theory, we have given the evidence of the non-Abelian dual Meissner effect
claimed by us, and found the chromoelectric flux tube by measuring the chromo
flux created by a static quark-antiquark pair. We have determined that the
type of vacuum for SU(3) case is of type I, which is in sharp contrast to the
$SU(2)$ case: the border of type I and type II.\cite{DMeisner-TypeI2013}.

In this talk, we focus on the confinement/deconfinement phase transition and
the non-Abelian dual superconductivity at finite temperature: We measure a
Polyakov loop average and correlation functions of the Polyakov loops which
are defined for both the original YM\ field and extracted $V$-field to examine
the $V$-field dominance in the Polyakov loop at finite temperature. Then, we
measure the chromoelectric flux between a pair of static quark and antiquark
at finite temperature, and investigate its relevance to the phase transition
and the non-Abelian dual Meissner effect.

\section{Method}

We introduce a new formulation of the lattice YM theory in the minimal option,
which extracts the dominant mode of the quark confinement for $SU(3)$ YM
theory\cite{abeliandomSU(3),lattice2010}, since we consider the quark
confinement in the fundamental representation. Let $U_{x,\mu}=X_{x,\mu
}V_{x,\mu}$ be the decomposition of YM link variable $U_{x,\mu}$, where
$V_{x,\mu}$ could be the dominant mode for quark confinement, and $X_{x,\mu}$
the remainder part. The YM field and the decomposed new variables are
transformed by full $SU(3)$ gauge transformation $\Omega_{x}$ such that
$V_{x,\mu}$ is transformed as the gauge link variable and $X_{x,\mu}$ as the
site variable:
\begin{subequations}
\label{eq:gaugeTransf}%
\begin{align}
U_{x,\mu} &  \longrightarrow U_{x,\nu}^{\prime}=\Omega_{x}U_{x,\mu}%
\Omega_{x+\mu}^{\dag},\\
V_{x,\mu} &  \longrightarrow V_{x,\nu}^{\prime}=\Omega_{x}V_{x,\mu}%
\Omega_{x+\mu}^{\dag},\text{ \ }X_{x,\mu}\longrightarrow X_{x,\nu}^{\prime
}=\Omega_{x}X_{x,\mu}\Omega_{x}^{\dag}.
\end{align}
The decomposition is given by solving the defining equation:
\end{subequations}
\begin{subequations}
\label{eq:DefEq}%
\begin{align}
&  D_{\mu}^{\epsilon}[V]\mathbf{h}_{x}:=\frac{1}{\epsilon}\left[  V_{x,\mu
}\mathbf{h}_{x+\mu}-\mathbf{h}_{x}V_{x,\mu}\right]  =0,\label{eq:def1}\\
&  g_{x}:=e^{i2\pi q/3}\exp(-ia_{x}^{0}\mathbf{h}_{x}-i\sum\nolimits_{j=1}%
^{3}a_{x}^{(j)}\mathbf{u}_{x}^{(j)})=1,\label{eq:def2}%
\end{align}
where $\mathbf{h}_{x}$ is an introduced color field $\mathbf{h}_{x}%
=\xi(\lambda^{8}/2)\xi^{\dag}$ $\in\lbrack SU(3)/U(2)]$ with $\lambda^{8}$
being the Gell-Mann matrix and $\xi$ an $SU(3)$ group element. The variable
$g_{x}$ is an undetermined parameter from Eq.(\ref{eq:def1}), $\mathbf{u}%
_{x}^{(j)}$ 's are $su(2)$-Lie algebra values, and $q_{x}$ an integer value
$\ 0,1,2$. These defining equations can be solved exactly \cite{exactdecomp},
and the solution is given by
\end{subequations}
\begin{subequations}
\label{eq:decomp}%
\begin{align}
X_{x,\mu} &  =\widehat{L}_{x,\mu}^{\dag}\det(\widehat{L}_{x,\mu})^{1/3}%
g_{x}^{-1},\text{ \ \ \ }V_{x,\mu}=X_{x,\mu}^{\dag}U_{x,\mu}=g_{x}\widehat
{L}_{x,\mu}U_{x,\mu},\\
\widehat{L}_{x,\mu} &  =\left(  L_{x,\mu}L_{x,\mu}^{\dag}\right)
^{-1/2}L_{x,\mu},\text{ \ \ \ \ \ \ }L_{x,\mu}=\frac{5}{3}\mathbf{1}+\frac
{2}{\sqrt{3}}(\mathbf{h}_{x}+U_{x,\mu}\mathbf{h}_{x+\mu}U_{x,\mu}^{\dag
})+8\mathbf{h}_{x}U_{x,\mu}\mathbf{h}_{x+\mu}U_{x,\mu}^{\dag}\text{ .}%
\end{align}
Note that the above defining equations correspond to the continuum version:
$D_{\mu}[\mathcal{V}]\mathbf{h}(x)=0$ and $\mathrm{tr}(\mathbf{h}%
(x)\mathcal{X}_{\mu}(x))$ $=0$, respectively. In the naive continuum limit, we
have the corresponding decomposition $\mathbf{A}_{\mathbf{\mu}}(x)=\mathbf{V}%
_{\mu}(x)+\mathbf{X}_{\mu}(x)$ in the continuum theory\cite{exactdecomp} as
\end{subequations}
\begin{subequations}
\begin{align}
\mathbf{V}_{\mu}(x) &  =\mathbf{A}_{\mathbf{\mu}}(x)-\frac{4}{3}\left[
\mathbf{h}(x),\left[  \mathbf{h}(x),\mathbf{A}_{\mathbf{\mu}}(x)\right]
\right]  -ig^{-1}\frac{4}{3}\left[  \partial_{\mu}\mathbf{h}(x),\mathbf{h}%
(x)\right]  ,\\
\mathbf{X}_{\mu}(x) &  =\frac{4}{3}\left[  \mathbf{h}(x),\left[
\mathbf{h}(x),\mathbf{A}_{\mathbf{\mu}}(x)\right]  \right]  +ig^{-1}\frac
{4}{3}\left[  \partial_{\mu}\mathbf{h}(x),\mathbf{h}(x)\right]  .
\end{align}

The decomposition is uniquely obtained as the solution (\ref{eq:decomp}) of
Eqs.(\ref{eq:DefEq}), if color fields$\{\mathbf{h}_{x}\}$ are obtained. To
determine the configuration of color fields, we use the reduction condition to
formulate the new theory written by new variables ($X_{x,\mu}$,$V_{x,\mu}$)
which is equipollent to the original YM theory. Here, we use the reduction
functional:
\end{subequations}
\begin{equation}
F_{\text{red}}[\mathbf{h}_{x}]=\sum_{x,\mu}\mathrm{tr}\left\{  (D_{\mu
}^{\epsilon}[U_{x,\mu}]\mathbf{h}_{x})^{\dag}(D_{\mu}^{\epsilon}[U_{x,\mu
}]\mathbf{h}_{x})\right\}  , \label{eq:reduction}%
\end{equation}
and then color fields $\left\{  \mathbf{h}_{x}\right\}  $ are obtained by
minimizing the functional (\ref{eq:reduction}).

\section{Result}

We generate YM gauge configurations$\{U_{x,\mu}\}$ at finite temperature using
the standard Wilson action on lattices $L^{3}\times N_{T}$: $L=24$,
$N_{T}=6,8,10,14,24$ with $\beta=6.0$, $L=24$,$\ N_{T}=4,6,8,10,12,14,24$ with
$\beta=6$.$2$ and $L=24$, $N_{T}=4,6$ with $\beta=6.4.$ The gauge link
decomposition $U_{x,\mu}=X_{x,\mu}V_{x,\mu}$ is obtained by the formula
(\ref{eq:decomp}) given in the previous section, after \ the color field
configuration $\{\mathbf{h}_{x}\}$ is obtained by solving the reduction
condition of minimizing the functional eq(\ref{eq:reduction}) for each gauge
configuration. In the measurement of the Polyakov loop and Wilson loop, we
apply the APE smearing technique to reduce noises \cite{APEsmear}%
.\begin{figure}[ptb]
\begin{center}
\includegraphics[
height=5.5cm, angle=270]
{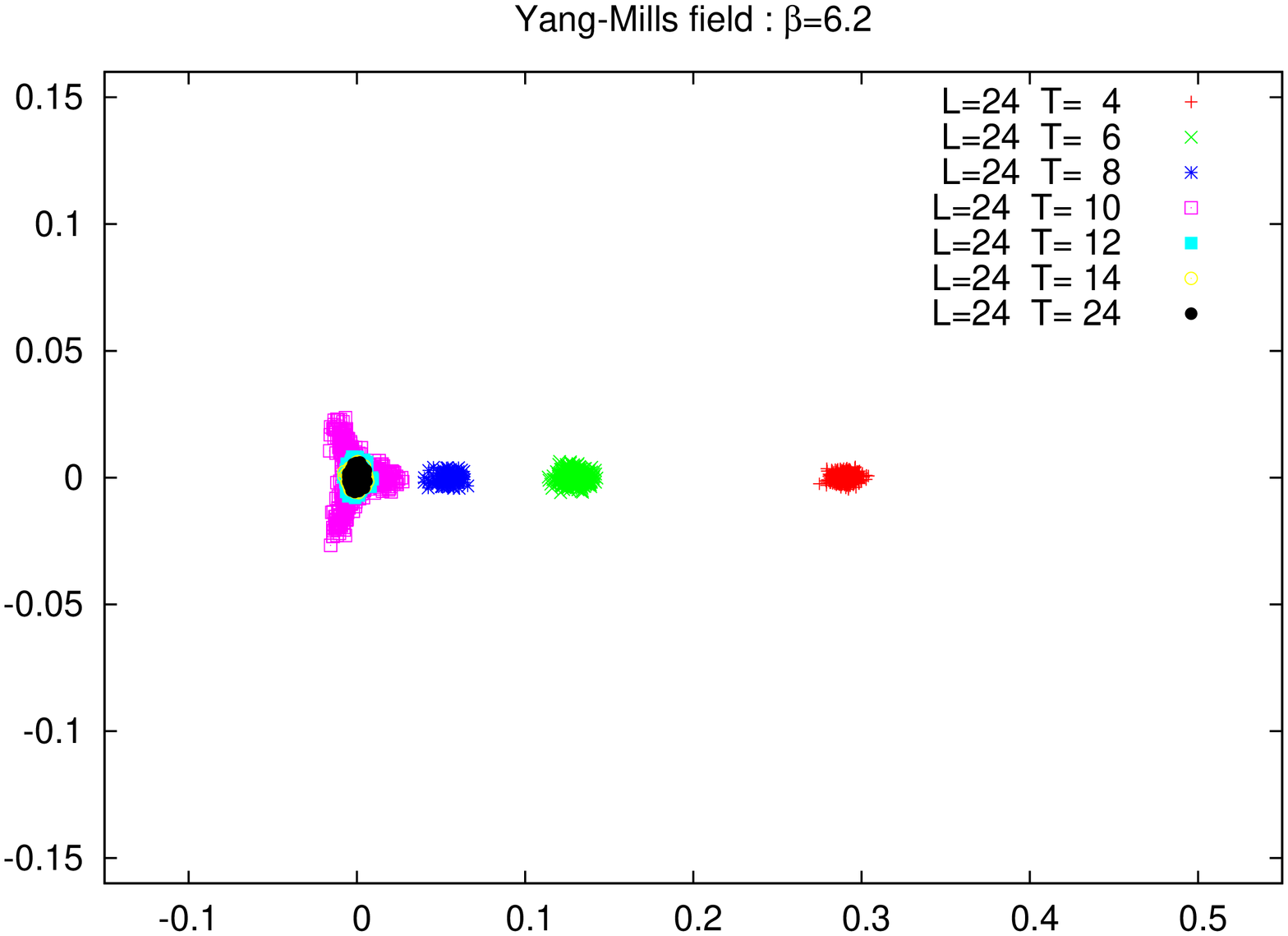} \ \ \includegraphics[
height=5.5cm, angle=270]
{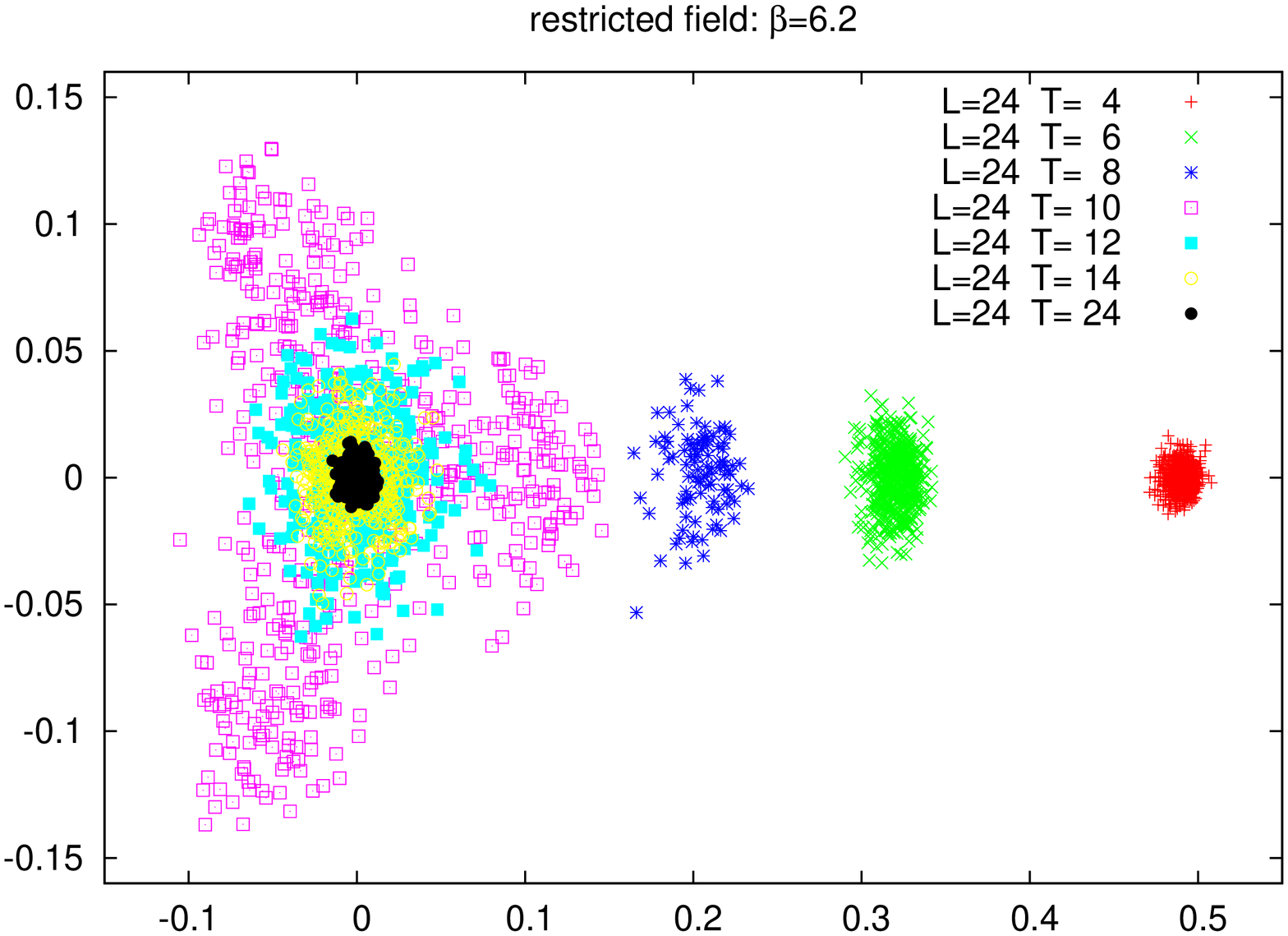}
\end{center}
\caption{{}The distribution of the space-averaged Polyakov loop for each
ocnfiguration, Eq.({\protect \ref{eq:PLP}}) (Left) For the YM field. (Right) For the
restricted field. }%
\label{Fig:PLP}%
\end{figure}

\begin{figure}[ptb]
\begin{center}
\includegraphics[
height=5.5cm, angle=270]
{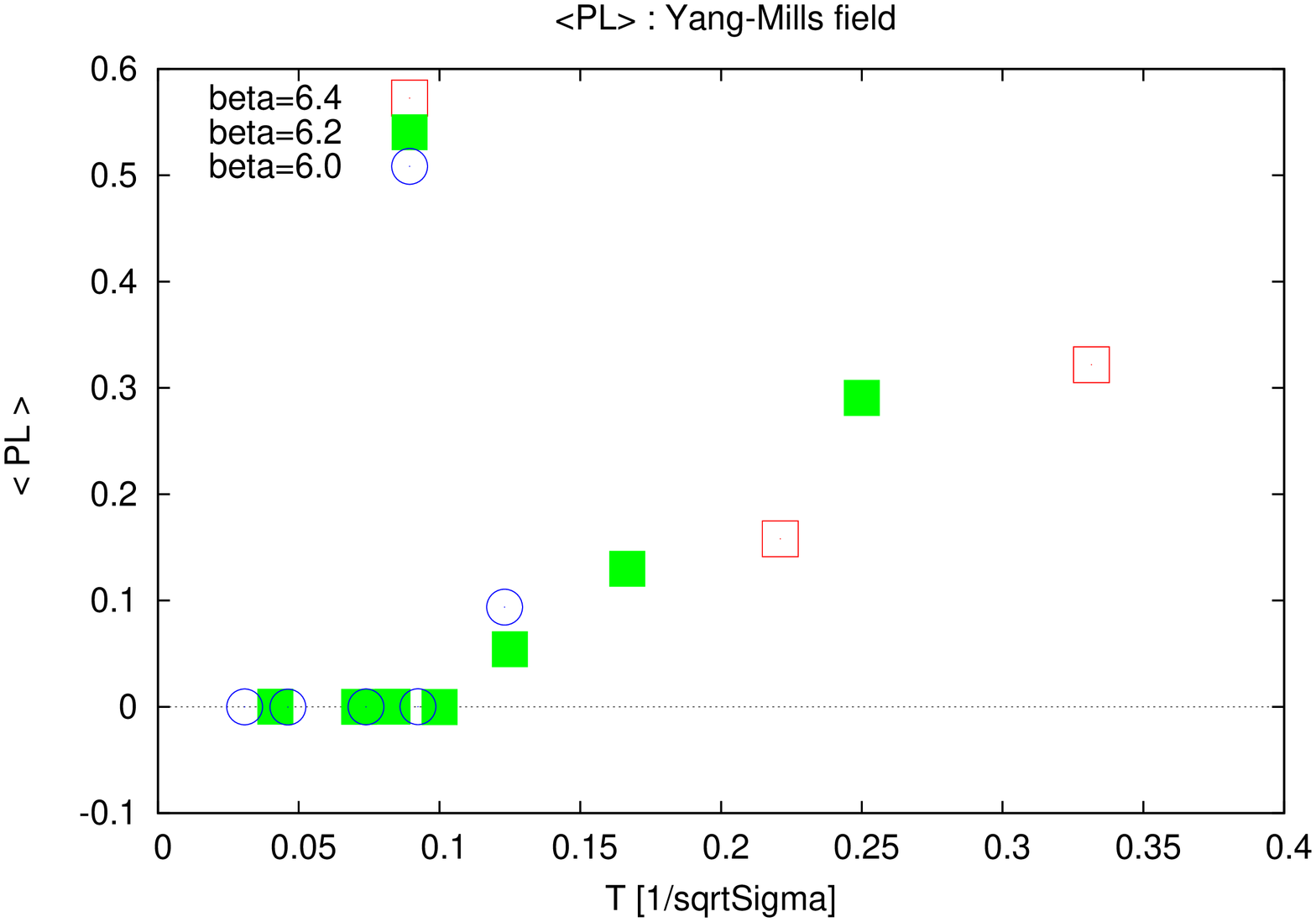} \ \ \includegraphics[
height=5.5cm, angle=270]
{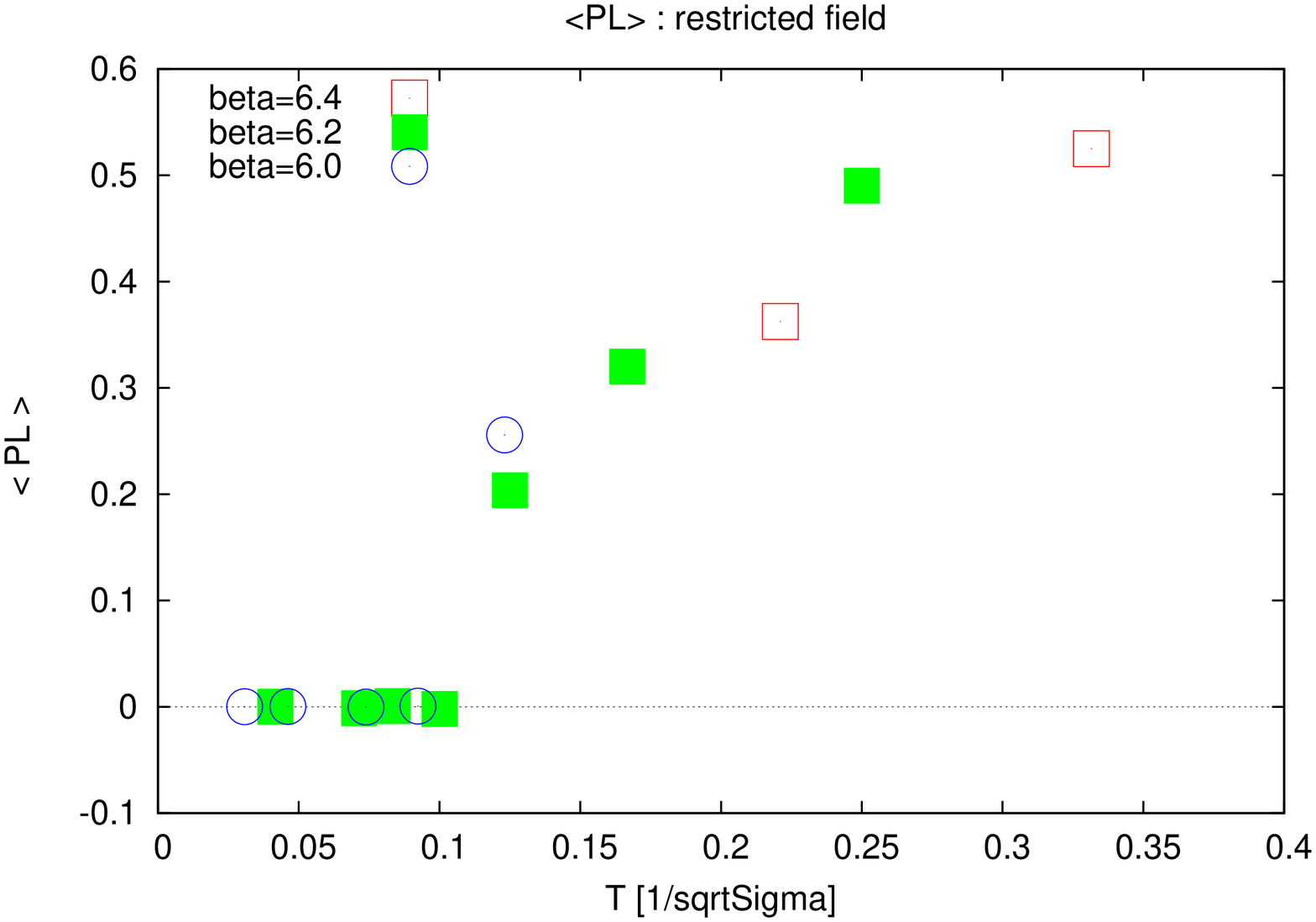}
\end{center}
\caption{{}The Polyakov loop average: (Left) For the YM field $\left\langle
P_{U}\right\rangle $ . (Right) For the restricted field $\left\langle
P_{V}\right\rangle .$}%
\label{fig:PLP-ave}%
\end{figure}

\begin{figure}[ptb]
\begin{center}
\includegraphics[
height=5.3cm, angle=270]
{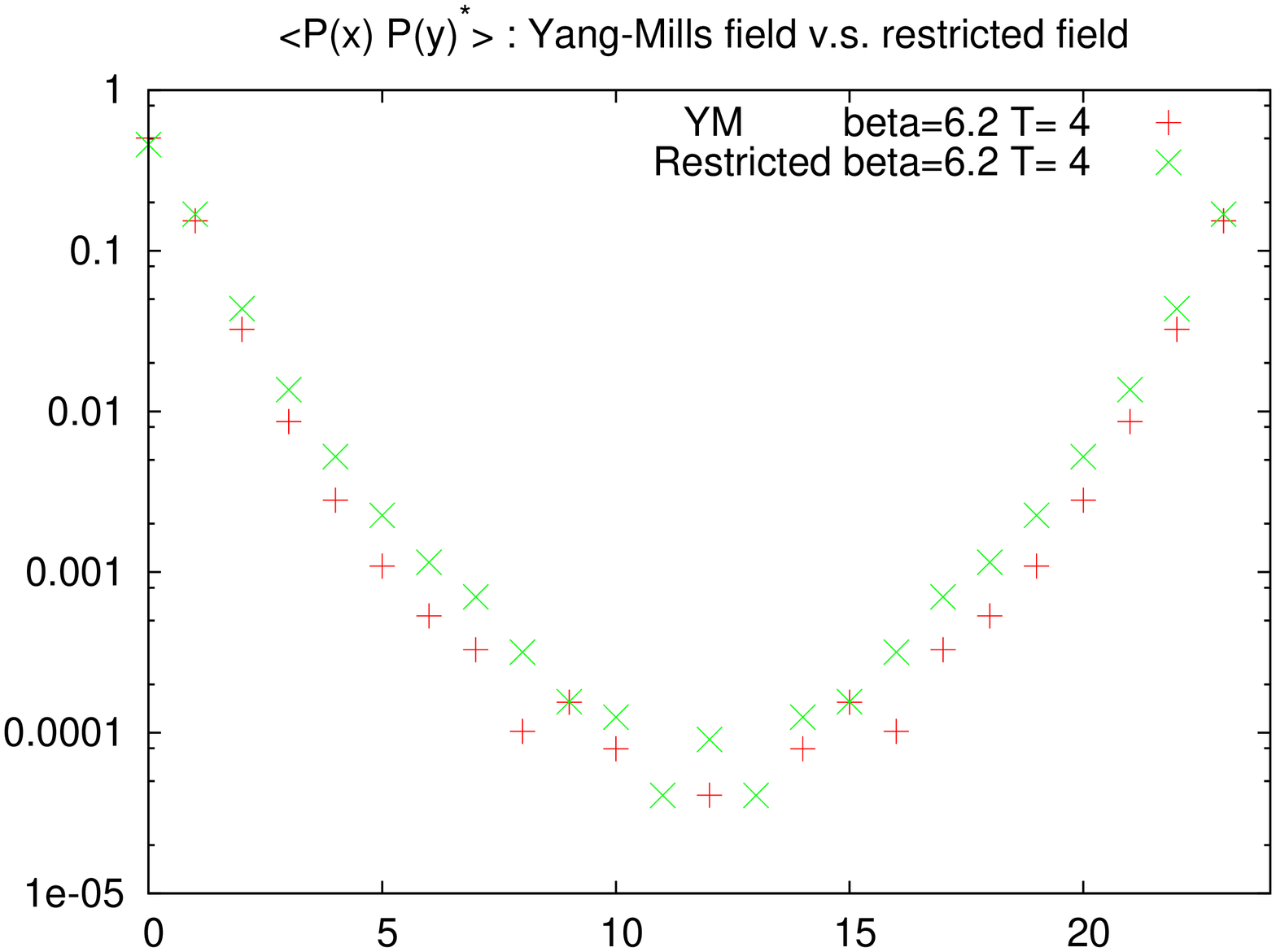} \includegraphics[
height=5.3cm, angle=270]
{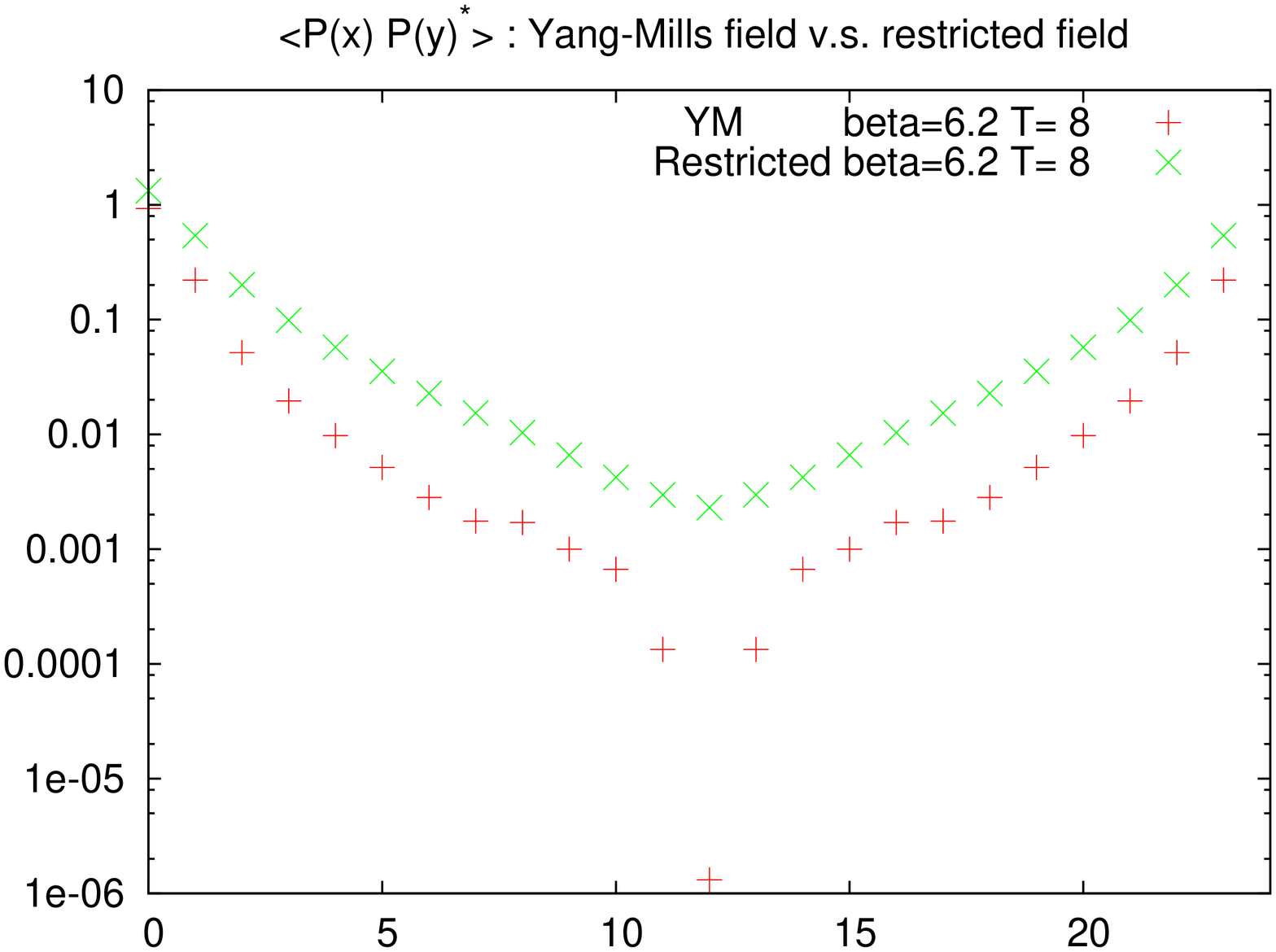} \includegraphics[
height=5.3cm, angle=270]
{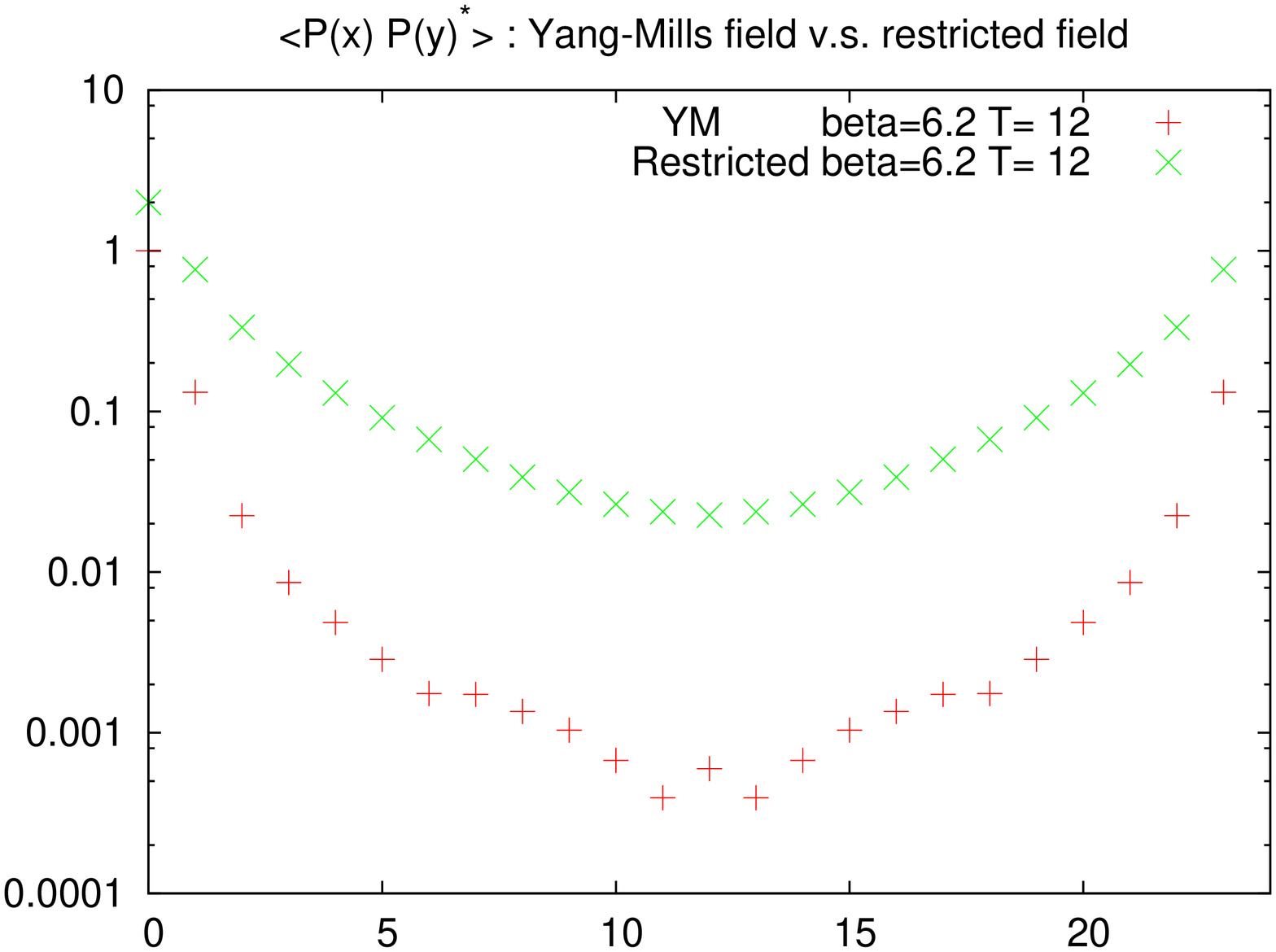} \ \
\end{center}
\caption{Comparison of \ the correlation function of the Polyakov loop for the
YM field and the restricted field. (Left)\ At high temperature. (Middle) Near
critical temperature. (Right) At low temperature.}%
\label{fig:PLP-correlations}%
\end{figure}

Figure \ref{Fig:PLP} show the distribution of space-averaged Polyakov loops
for each configurations:
\begin{equation}
P_{U}:=L^{-3}\sum\mathrm{tr}\left(  \prod\nolimits_{t=1}^{N_{T}}U_{(\vec
{x},t),4}\right)  ,\text{ \ \ \ }P_{V}:=L^{-3}\sum\mathrm{tr}\left(
\prod\nolimits_{t=1}^{N_{T}}V_{(\vec{x},t),4}\right)  .\label{eq:PLP}%
\end{equation}
The left panel shows the distribution of the YM\ field and the right panel
that of the restricted field ($V$-field). Then, we obtain the Polyakov loop
average for configurations, which is the conventional order parameter for
confinement and deconfinement phase transition in $SU(3)$ YM\ theory. Figure
\ref{fig:PLP-ave} shows the Polyakov loop average for the YM field
$\left\langle P_{U}\right\rangle $ (left panel) and restricted field
$\left\langle P_{V}\right\rangle $ (right panel). Each panel shows the same
critical temperature of confinement/deconfinement phase transition.
Then, we investigate two-point correlation function of the
Polyakov loop:%
\begin{equation}
D_{U}(x-y):=\left\langle P_{U}(x)^{\ast}P_{U}(y)\right\rangle -\left\langle
|P_{U}|^{2}\right\rangle ,\text{ \ \ \ }D_{V}(x-y):=\left\langle
P_{V}(x)^{\ast}P_{V}(y)\right\rangle -\left\langle |P_{V}|^{2}\right\rangle ,
\end{equation}
Figure \ref{fig:PLP-correlations} shows the comparison of the $D_{U}(x-y)$ and
$D_{V}(x-y)$ each temperature. Every panel shows that the YM-field and
restricted field ($V$-field) have the same profile, i.e., we can extract the
dominant mode for the quark confinement by the $V$-field.\begin{figure}[ptb]
\begin{center}
\includegraphics[
height=4.5cm]
{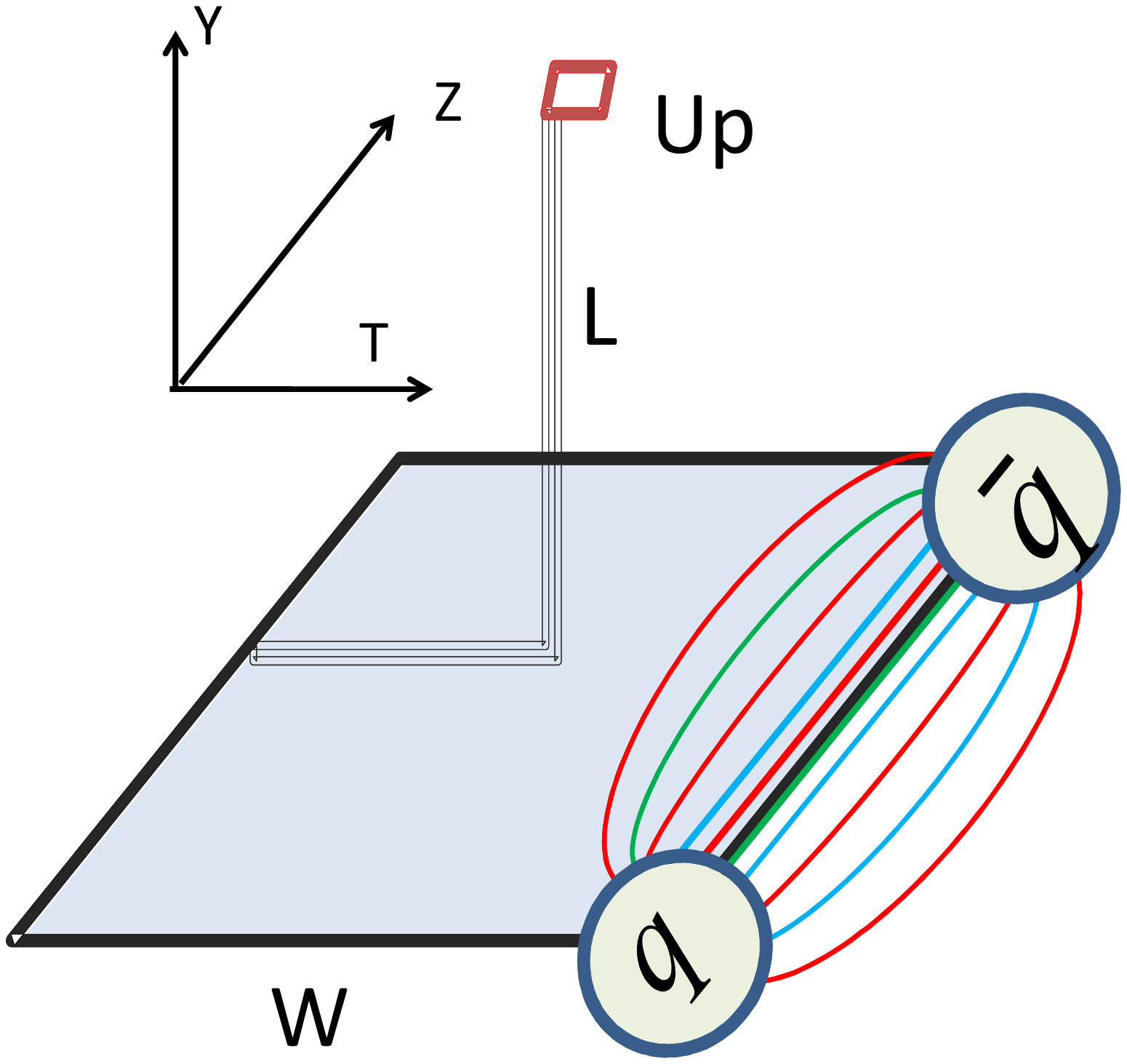} \ \ \ \ \ \ \ \includegraphics[
height=3.8cm]
{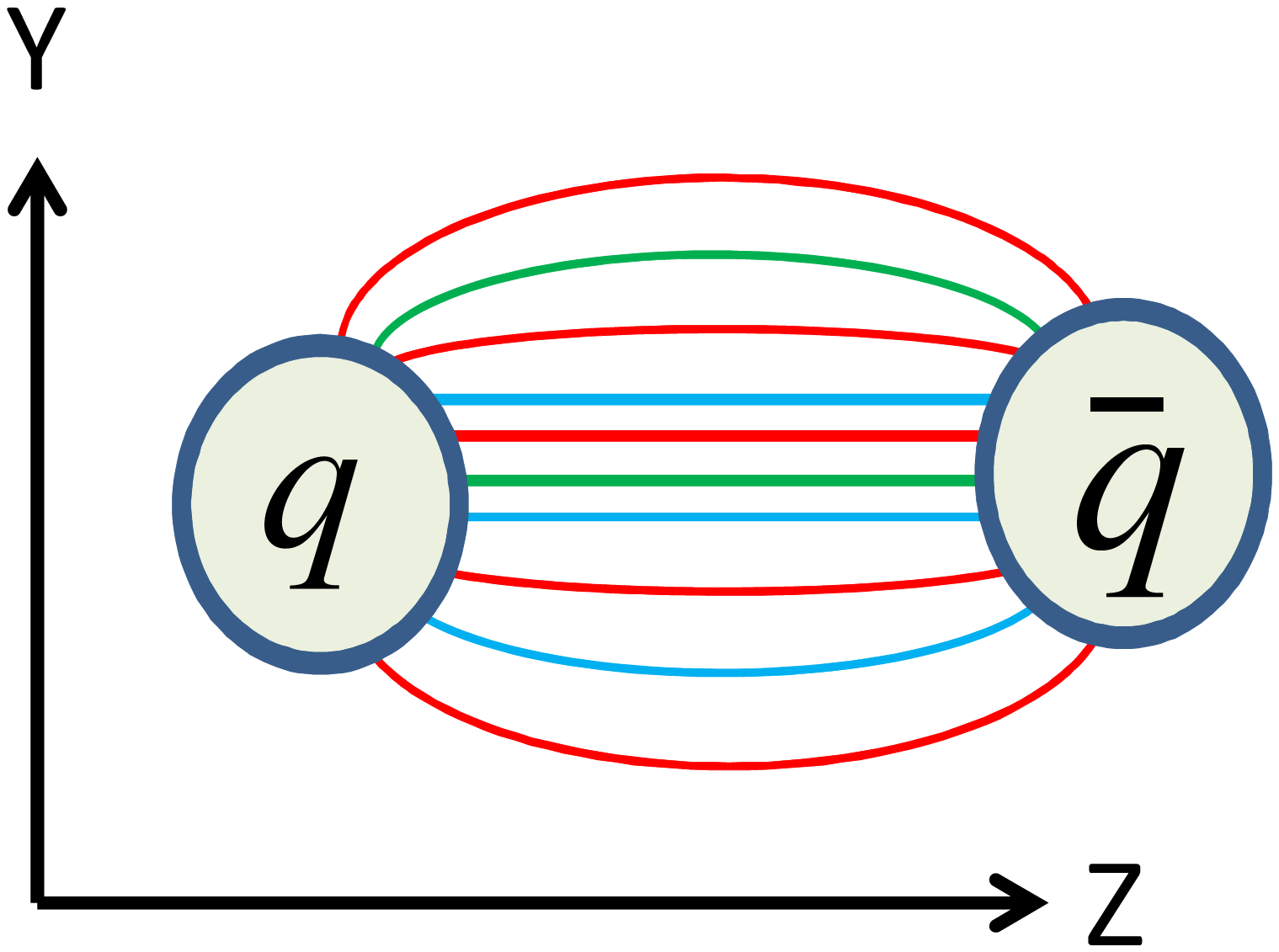}
\end{center}
\caption{(Left) The connected correlator ($U_{p}LWL^{\dag})$ between a
plaquette $U_{p}$ and the Wilson loop $W$. (Right) Measurement of  the
chromo-flux in the Y-Z plane. }%
\label{fig:measure}%
\end{figure}

Next, we investigate the non-Abelian dual Meissner effect at finite
temperature. To investigate the chromo flux, we use the gauge invariant
correlation function which is used at zero temperature. The chromo flux
created by a quark-antiquark pair is measured by using a gauge-invariant
connected correlator of the Wilson loop \cite{Giacomo}:%
\begin{equation}
\rho_{W}:=\frac{\left\langle \mathrm{tr}\left(  U_{p}L^{\dag}WL\right)
\right\rangle }{\left\langle \mathrm{tr}\left(  W\right)  \right\rangle
}-\frac{1}{3}\frac{\left\langle \mathrm{tr}\left(  U_{p}\right)
\mathrm{tr}\left(  W\right)  \right\rangle }{\left\langle \mathrm{tr}\left(
W\right)  \right\rangle },\label{eq:Op}%
\end{equation}
where $W$ represents a quark-antiquark pair settled by the Wilson loop in Z-T
plane, $U_{p}$ a plaquette variable as the probe operator for measuring the
field strength, and $L$ the Wilson line connecting the source $W$ and the
probe $U_{p}.$ (see the left panel\ of Figure \ref{fig:measure}). The symbol
$\left\langle \mathcal{O}\right\rangle $ denotes the average of the operator
$\mathcal{O}$ over the space and the ensemble of the configurations. Note that
this is sensitive to the field strength rather than the disconnected one.
Indeed, in the naive continuum limit, the connected correlator $\rho_{W}$ is
given by $\ \rho_{W}\overset{\varepsilon\rightarrow0}{\simeq}g\epsilon
^{2}\left\langle \mathcal{F}_{\mu\nu}\right\rangle _{q\bar{q}}:=\frac
{\left\langle \mathrm{tr}\left(  g\epsilon^{2}\mathcal{F}_{\mu\nu}L^{\dag
}WL\right)  \right\rangle }{\left\langle \mathrm{tr}\left(  W\right)
\right\rangle }+O(\epsilon^{4})$. Thus, the chromo field strength is given by
$\ F_{\mu\nu}=\sqrt{\frac{\beta}{6}}\rho_{W}$. Note that at finite temperature
we must use the operator with the same size in the temporal direction, and the
quark and antiquark pair is replaced by a pair of the Polyakov loop with the
opposite direction.\begin{figure}[ptb]
\begin{center}
\includegraphics[
height=6cm, angle=270]
{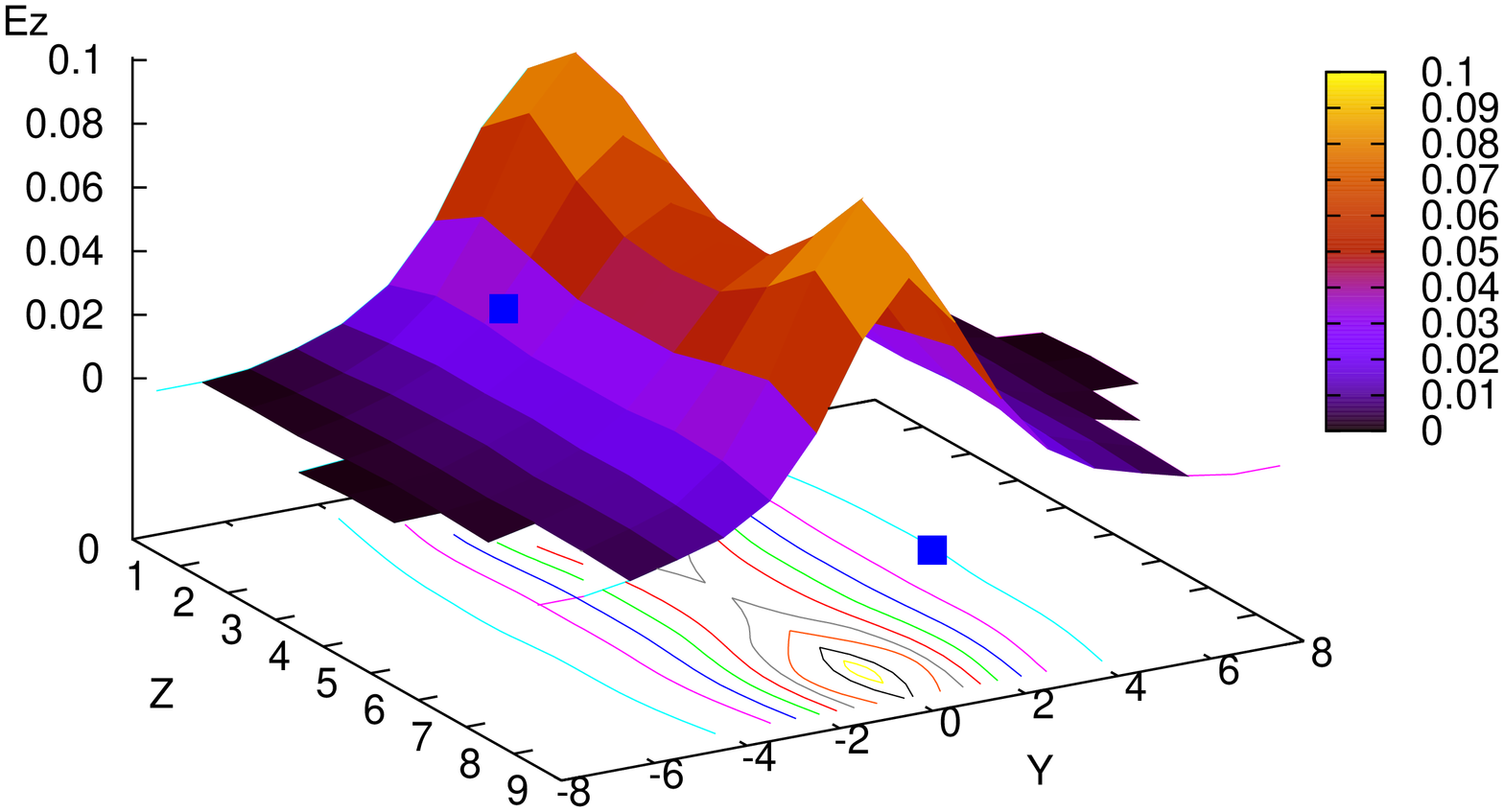} \ \ \ \ \ \ \ \includegraphics[
height=6cm, angle=270]
{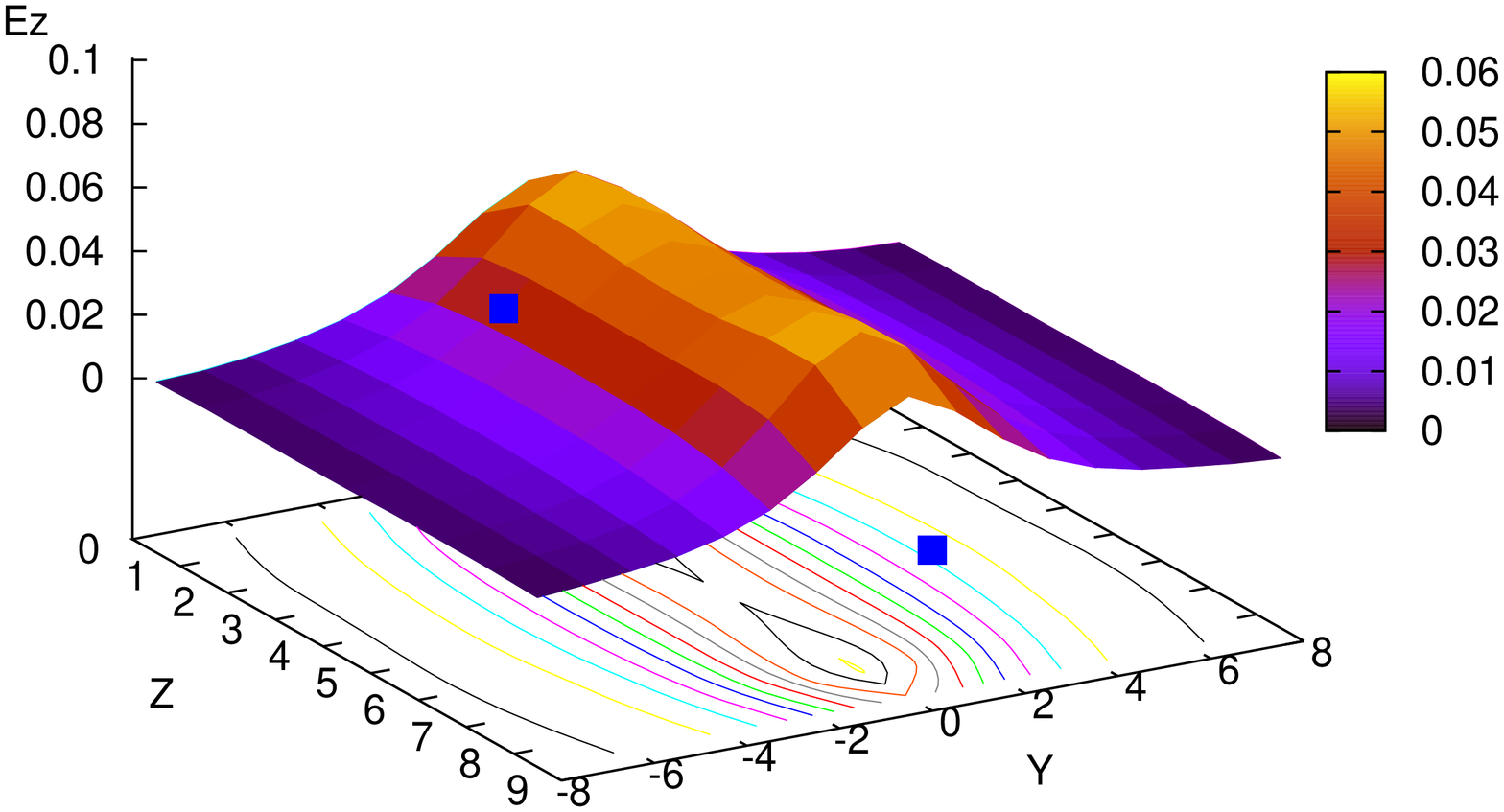}
\end{center}
\caption{{}The distribution in $Y$-$Z$ plane of the chromoelectric field
$E_{z}$ connecting a pair of quark and antiquark: (Left panel) chromoelectric
field produced from the original YM field, (Right panel) chromoelectric field
produced from the restricted field. }%
\label{fig:Ez-T0}%
\end{figure}\begin{figure}[ptb]
\begin{center}
\includegraphics[
height=6cm, angle=270]
{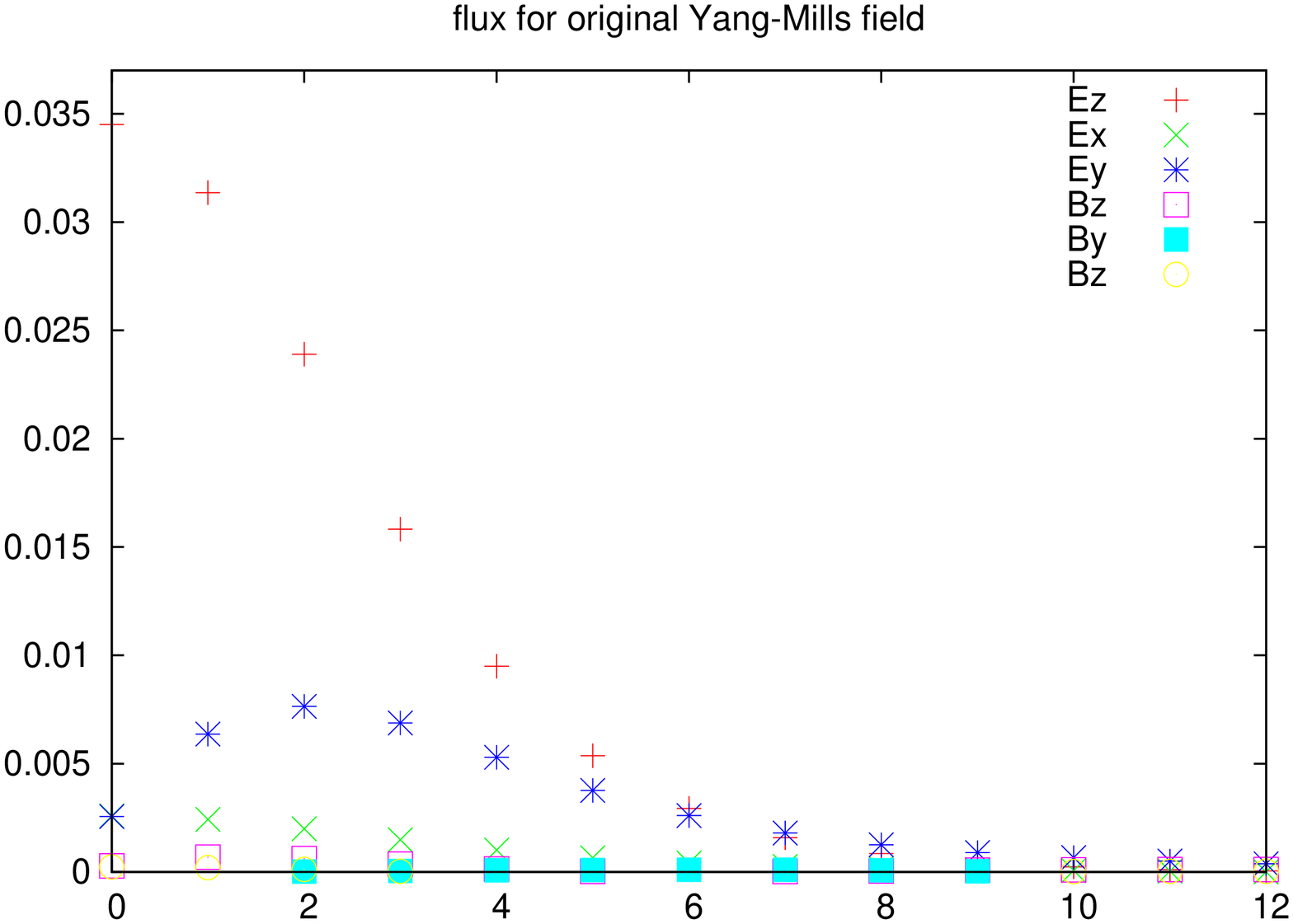} \ \ \ \ \ \ \includegraphics[
height=6cm, angle=270]
{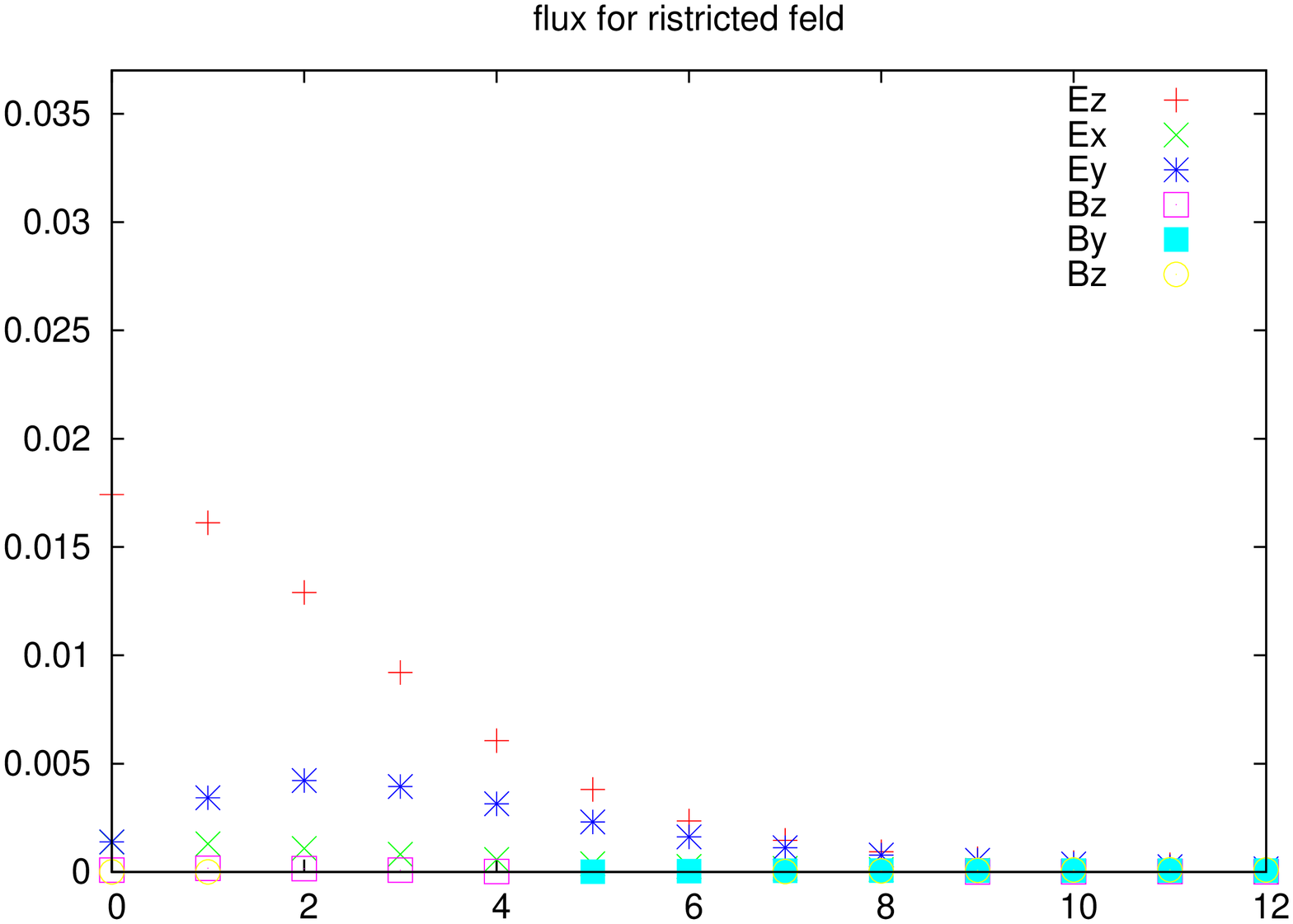}
\end{center}
\caption{{}The chromo-flux created by the quark-antiquark pair in the plane
$z=1/3R$ for a quark at $z=0$ and an antiquark at $z=R$ (Fig.{\protect \ref{fig:measure}}%
) by moving the probe, $U_{p}$ or $V_{p}$ along the y-direction. (Left) For
the YM-field. (Right) For the restricted field.}%
\label{fig:flux-T}%
\end{figure}

Figure \ref{fig:Ez-T0} shows the measurement of the Z-component $E_{z}$ of
chromoelectric flux at zero temperature. We observe the chromoelectric flux
tube only in the direction connecting quark and antiquark pair, while the
other components take vanishing values \cite{DMeisner-TypeI2013}. Figure
\ref{fig:flux-T} shows the measurement of chromoelectric and chromomagnetic
flux at high temperature $T>T_{c}$ $\ $(for the lattice $N_{T}=6,$ $\beta
=6.2$\.{)}. We measure the chromo-flux of quark-antiquark pair in the plane
$z=1/3R$ for a quark at $z=0$ and an antiquark at $z=R$ (Fig.\ref{fig:measure}%
) by moving the probe, $U_{p}$ or $V_{p}$ along the y-direction. We can
observe no more squeezing of the chromoelectric flux tube, but non-vanishing
$E_{y}$ component \ in the chromoelectric flux. This shows the disappearance
of the dual Meisner effect at high temperature.

\section{Summary and outlook}

We have studied the dual superconductivity for $SU(3)$ YM theory by using our
new formulation of YM theory on a lattice. We have extracted the restricted
field ($V$-field) from the YM field which plays a dominant role in confinement
of quark (fermion in the fundamental representation) at finite temperature,
i.e., the restricted field dominance in Polyakov loop. Then we have measured
the chromoelectric and chromomagnetic flux for both the original YM field and
the restricted field at low temperature. We have observed evidences of the
dual Messier effect of $SU(3)$ YM theory, i.e., the chromoelectric flux tube
and the associated non-Abelian magnetic monopoles created by quark and
antiquark pair. At high temperature ($T>T_{c}$) in the deconfinement phase, we
have observed the disappearance of the dual Meissner effect by measuring the
chromo flux. \ Note that, the Polyakov loop average cannot be the direct
signal of the dual Meissner effect or magnetic monopole condensation.
Therefore, it is important to find a order parameter which detect the dual
Meissner effect directly, and to investigate whether or not the order
parameter in view of the dual Meissner effect gives the same critical
temperature as that of the Polyakov loop average.

\subsection*{Acknowledgement}

This work is supported by Grant-in-Aid for Scientific Research (C) 24540252
from Japan Society for the Promotion Science (JSPS), and also in part by JSPS
Grant-in-Aid for Scientific Research (S) 22224003. The numerical calculations
are supported by the Large Scale Simulation Program No.12-13 (FY2012), No.
12/13-20 (FY2012/13) of High Energy Accelerator Research Organization (KEK).

\end{document}